\documentclass[a4paper]{jpconf}
\usepackage{graphicx,hyperref,iopams}
\usepackage{ulem}

\usepackage{xcolor}

\begin{document}
	\title{Internal Heating in Magnetars: Role of Electron Captures}
	
	\author{Nicolas Chamel$^{1,}$, Anthea Francesca Fantina $^{2,1}$, Lami Suleiman $^{3,4}$, Julian-Leszek Zdunik $^{3}$	and Pawel Haensel $^{3}$}
	
	\address{$^{1}$Institute of Astronomy and Astrophysics,  Universit\'e Libre de Bruxelles,  CP 226, Boulevard du Triomphe, B-1050 Brussels, Belgium \\ 
		$^{2}$Grand Acc\'el\'erateur National d'Ions Lourds (GANIL), CEA/DRF-CNRS/IN2P3,  Boulevard Henri Becquerel, 14076 Caen, France \\ 
		$^{3}$N. Copernicus Astronomical Center, Polish Academy of Sciences, Bartycka 18, PL-00-716 Warszawa, Poland \\ 
		$^{4}$Laboratoire Univers et Théories, Observatoire de Paris, CNRS, Université Paris Cité, 92195 Meudon, France}

	\ead{nicolas.chamel@ulb.be}
	
	\begin{abstract}
The role of electron captures by nuclei in the shallow heating of magnetars is further investigated using both nuclear measurements and the theoretical atomic mass table HFB-27. Starting from the composition of the outer crust in full equilibrium, we have calculated  the onset of electron captures and the heat released due to the slow  decay of the magnetic field. 
Numerical results are found to be similar to those previously obtained  with the HFB-24 atomic mass model  and are  consistent with  neutron-star cooling data. 
	\end{abstract}
	
	\section{Introduction}
	
	With their extreme magnetic fields, soft gamma-ray repeaters and anomalous X-ray pulsars are very active types of neutron stars exhibiting outbursts and more rarely giant flares releasing energies up to $\sim 10^{46}$~erg within a second (see e.g. \cite{esposito2021} for a recent review). So far, 24 of these magnetars have been identified according to the McGill Online Magnetar Catalog\footnote{\url{http://www.physics.mcgill.ca/~pulsar/magnetar/main.html}} \cite{mcgill2014}. Their magnetic field is also thought to power their persistent X-ray luminosity $\sim 10^{33}-10^{35}$ erg/s, which is well in excess of their rotational energy and which implies 
	higher surface temperatures than in weakly magnetized neutron stars of the same age~\cite{vigano2013}. 
	The most widely accepted heating mechanism involves crustal deformations beyond the elastic limit induced by magnetic stresses (see, e.g., Ref.~\cite{degrandis2020}; see Ref.~\cite{beloborodov2016} for alternative models). Because the melting temperature of an electron-ion Coulomb solid increases with density~\cite{fantina2020,carreau2020}, this mechanism is most effective in the inner region of the crust. However, heat sources are not expected to be found 
	in such deep region of the star~\cite{kaminker2006,kaminker2009}. 
	
	Alternatively, the magnetic energy may be converted into heat through electron captures by nuclei in the outer crust~\cite{cooper2010}. This mechanism is analogous to crustal heating in accreting neutron stars~\cite{haensel1990}, the matter compression being induced here by the magnetic-field decay rather than accretion from a stellar companion. We have recently estimated the maximum amount of heat that could be possibly released by these processes and the location of the heat sources~\cite{chamel2021}. In addition to electron captures, we have also considered pycnonuclear fusion of light elements. After briefly reviewing our assumptions and our treatment of dense matter, new results are reported using a different nuclear model.

	\section{Compression-induced reactions in magnetar crusts}
	
	\subsection{Initial composition}
	
	We assume that each crustal layer is made of nuclei $(A,Z)$ with proton number $Z$ and mass number $A$ in a charge neutralizing background of free electrons. 
	
	We take into account Landau-Rabi quantization for electrons (see, e.g. Chapter 4 of Ref.~\cite{haensel2007}) and  consider that electrons all lie in the lowest level, which occurs whenever $B\gtrsim 5.72\times 10^{16}$~G~\cite{chamel2015b}. Quantization effects are expected to be very strong for such fields because the temperatures prevailing in a magnetar, $T\sim 10^8-10^9$~K, are much lower than the characteristic temperature 
	\begin{equation}\label{eq:TB}
		T_B=\frac{m_e c^2}{k_B} B_\star\approx 5.93\times 10^9 B_\star~\rm K\, , 
	\end{equation}
	where $k_B$ denotes Boltzmann's constant, $c$ is the speed of light, $m_e$ is the electron mass, and we have introduced the dimensionless magnetic field strength $B_\star\equiv B/B_{\rm rel}\gg1 $ with: 
	\begin{equation}
		\label{eq:Bcrit}
		B_{\rm rel}=\frac{m_e^2 c^3}{e \hbar}\approx 4.41\times 10^{13}\, \rm G\, , 
	\end{equation}
	($e$ is the elementary electric charge and $\hbar$ is the Planck-Dirac constant). 
	
	The electron-ion plasma may not necessarily be in a solid state, especially in the shallow layers. A lower bound on the melting temperature can be found by ignoring the magnetic field~\cite{potekhin2013}, and is typically of order $T_m\sim 10^9$~K (see Ref.~\cite{fantina2020} for more details). 	

In this study, thermal effects on thermodynamic quantities will be neglected. 
	
	Assuming the crust is initially in full thermodynamic equilibrium in presence of some magnetic field, the composition is found by minimizing the Gibbs free energy per nucleon (see Ref.~\cite{chamelfantina2015}). 
	This minimization can be performed numerically in a very efficient way by using the iterative approach proposed in Ref.~\cite{chamel2020}.

	\subsection{Magnetic field decay and electron captures} 
	
	The compression of the crust induced by the decay of the magnetic field occurs very slowly, on a typical time scale of millions of years~\cite{pons2019}. The capture of one electron by the nucleus $(A,Z)$ (in its ground state since matter is assumed to be initially in equilibrium) becomes allowed once the local pressure attains some threshold value $P_\beta(A,Z,B_\star)$, which can be calculated analytically and is given by~\cite{chamel2021}
	\begin{eqnarray}\label{eq:exact-Pbeta}
		P_{\beta}(A,Z,B_\star)&\approx& \frac{B_\star m_e c^2 }{4 \pi^2 \lambda_e^3 }\biggl[\gamma_e\sqrt{\gamma_e^2-1}-\ln\left(\sqrt{\gamma_e^2-1}+\gamma_e\right) \nonumber \\ 
		&&+\frac{C \alpha }{3}\left(\frac{4 B_\star Z^2 }{\pi^2}\right)^{1/3} \left(\gamma_e^2-1\right)^{2/3} \biggr] \, ,
	\end{eqnarray} 
	where $\lambda_e=\hbar/(m_e c)$ is the electron Compton wavelength, $\alpha=e^2/(\hbar c)$ is the fine-structure constant, $C$ is a dimensionless constant characterizing the spatial arrangements of nuclei and for which we adopt the Wigner-Seitz value $C=-9/10 (4\pi/3)^{1/3}\approx -1.4508$~\cite{salpeter1954}. Here $\gamma_e$ is the threshold electron Fermi energy in units of $m_e c^2$ given by \cite{chamel2021}
	\begin{equation}\label{eq:exact-gammae}
		\gamma_e=\left\{ \begin{array}{ll}
			\displaystyle 8|\bar F(Z, B_\star)|^{3/2}\, {\rm cosh}^3\left(\frac{1}{3}{\rm arccosh~} \frac{\gamma_e^{\beta}}{2 |\bar F(Z, B_\star)|^{3/2}}\right) & {\rm if} \ \upsilon\geq 1\, ,\\
			\displaystyle 8|\bar F(Z, B_\star)|^{3/2}\, \cos^3\left( \frac{1}{3}\arccos \frac{\gamma_e^{\beta}}{2 |\bar F(Z, B_\star)|^{3/2}}\right) & {\rm if} \ 0\leq \upsilon< 1\, ,
		\end{array}\right.
	\end{equation}
	with 
	\begin{equation}
		\bar F(Z,B_\star)\equiv \frac{1}{3} C\alpha \left(\frac{B_\star}{2\pi^2}\right)^{1/3} \biggl[Z^{5/3}-(Z-1)^{5/3} + \frac{1}{3} Z^{2/3}\biggr]\, ,
	\end{equation}
	\begin{equation}\label{eq:muebeta}
		\gamma_e^{\beta}\equiv -\frac{Q_{\rm EC}}{m_e c^2} + 1 \, ,
	\end{equation}
\begin{equation}
Q_{\rm EC}(A,Z) = M^\prime(A,Z)c^2-M^\prime(A,Z-1)c^2-E_{\rm ex}(A,Z-1)\, .
\end{equation}	
Here $M^\prime(A,Z)$ denotes the nuclear mass including the rest mass of $Z$ electrons and $E_{\rm ex}(A,Z-1)$ is the excitation energy of the daughter nuclei. The  mean baryon number density at the onset of the first electron capture is given by~\cite{chamel2021}
	\begin{equation}\label{eq:exact-rhobeta}
		n_\beta(A,Z,B_\star) = \frac{B_\star}{2\pi^2 \lambda_e^3} \frac{A}{Z} \sqrt{\gamma_e^2-1} \, . 
	\end{equation}
	
No heat is released by this first electron capture since it proceeds in quasiequilibrium. However, the daughter nucleus (in some excited state)  $(A,Z-1)$ is almost always unstable against a second electron capture off-equilibrium thus depositing some heat at the \emph{same} pressure $P_\beta(A,Z,B_\star)$. The maximum amount of heat per nucleus can be estimated analytically ignoring the fraction of energy carried away by neutrinos~\cite{chamel2021}
	\begin{eqnarray}
		\label{eq:heat-ocrust}
		\mathcal{Q}(A,Z,B_\star) &=&Q_{\rm EC}(A,Z-1)-Q_{\rm EC}(A,Z)+ 2 E_{\rm ex}(A,Z-1)  \nonumber \\
		&-& m_e c^2 C \alpha \left(\frac{B_\star }{2\pi^2}\right)^{1/3}(\gamma_e^2-1)^{1/6}\biggl[Z^{5/3}+(Z-2)^{5/3}-2 (Z-1)^{5/3}\biggr]\, .
	\end{eqnarray}
	Leaving aside the small terms on the second line, the maximum heat released is thus completely determined by nuclear data alone, independently of the magnetic field and whether the electron-ion plasma is solid or not.

	In the densest regions of the outer crust, fusion of the lightest nuclei might occur~\cite{yakovlev2006}. Considering that such reactions become allowed at pressure $P_{\rm pyc}=P_\beta$ yields the following upper bound on the heat released per parent nucleus: 
	\begin{eqnarray}
		\mathcal{Q}_{\rm pyc}(A,Z,B_\star)&=& M^\prime(A,Z-1)c^2-\frac{M^\prime(2A,2Z-2)c^2}{2}+E_{\rm ex}(A,Z-1)\nonumber \\
		&+& m_e c^2 C \alpha (1-2^{2/3}) \left(\frac{B_\star \gamma_e}{2 \pi^2} \right)^{1/3} (Z-1)^{5/3} \, ,
	\end{eqnarray}
	with $\gamma_e$ given by Eq.~(\ref{eq:exact-gammae}). 
	
	The bottom of the outer crust is marked by the emission of free neutrons~\cite{chamel2015b,chamel2015a}. We have not considered reactions occurring in deeper layers (inner crust).

	\section{Results and discussion} 
	
	\subsection{Numerical estimates}
	
\begin{figure}[h]
		\includegraphics{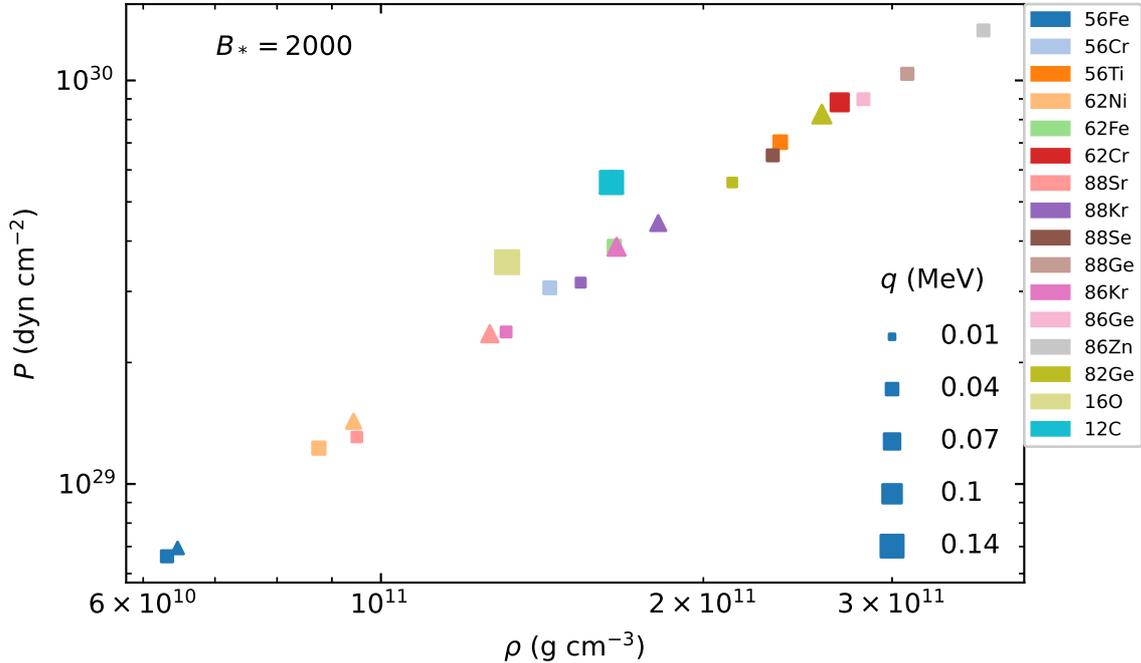}
		\caption{\label{fig:heat}Heat $q$ per nucleon (in MeV) from selected electron captures in the crust of a magnetar with a magnetic field $B_\star=2000$ in a pressure $P$ (in dyn~cm$^{-2}$)--mass density $\rho$ (in g cm$^{-3}$) diagram, considering transitions from the ground state of the parent nucleus to either the ground state (squares) or the first excited state (triangles) of the daughter nucleus. The size of each symbol is proportional to the amount of heat deposited as indicated. }
	\end{figure}

	We have calculated the crustal heat using nuclear masses and $Q_\beta(A,Z-1) = -Q_{\rm EC}(A,Z)$ values from the 2016 Atomic Mass Evaluation \cite{ame2016}, excitation energies from the Nuclear Data section of the International Atomic Energy Agency website\footnote{\url{https://www-nds.iaea.org/relnsd/NdsEnsdf/QueryForm.html}}, as well as 
predictions from the atomic  mass model HFB-27 based on self-consistent deformed Hartree-Fock-Bogoliubov calculations~\cite{goriely2013b}. Further details can be found in Ref.~\cite{chamel2021}. 
The magnetic field strength was set to $B_\star=2000$. The initial equilibrium composition was taken from Ref.~\cite{chamel2020}. As for possible contamination of the crust with light elements from the interstellar medium, we considered carbon and oxygen. 
	Results are summarized in Fig.~\ref{fig:heat}.  
	
	Comparing the present results with those previously obtained using the HFB-24 mass model in Ref.~\cite{chamel2021}, we notice that the reactions releasing the largest amount of heat are the same, namely, $^{12}\mathrm{C} \rightarrow {}^{12}\mathrm{Be}$, ${}^{16}\mathrm{O} \rightarrow {}^{16}\mathrm{C}$, and $^{62}\mathrm{Cr} \rightarrow {}^{62}\mathrm{Ti}$ considering ground-state-to-ground-state transitions (the heat released being 
	$\sim 0.14$~MeV, $\sim 0.15$~MeV, and $\sim 0.09$~MeV per nucleon), and $^{82} \mathrm{Ge} \rightarrow {}^{82}\mathrm{Zn}$, $^{88} \mathrm{Sr} \rightarrow {}^{88}\mathrm{Kr}$, $^{88} \mathrm{Kr} \rightarrow ^{88} \mathrm{Se}$, and $^{86} \mathrm{Kr} \rightarrow {}^{86}\mathrm{Se}$ considering ground-state-to-excited-state transitions (the heat released being $\sim 0.1$~MeV, $\sim 0.08$~MeV, $\sim 0.07$~MeV, $\sim 0.09$~MeV per nucleon, respectively).
	Note that for the latter reactions, as well as for the reactions involving carbon and oxygen, the amount of heat released is only determined by experimental data, thus independent of the adopted mass model (see also Tables~1 and 3 in Ref.~\cite{chamel2021}).
	The most noticeable difference between the predictions of the mass models HFB-24 and HFB-27 concerns the heat released in the ground-state-to-ground-state transitions $^{56}\mathrm{Ti} \rightarrow {}^{56}\mathrm{Ca}$ and ${}^{80}\mathrm{Zn} \rightarrow {}^{80}\mathrm{Ni}$.
	In the former (latter) reaction, the heat predicted by HFB-27 is a factor of $\sim 2$ ($\sim 3$) higher than that calculated using the HFB-24 mass model.
	The reason stems from the discrepancy in the $Q_{\rm EC}$ values, see Eq.~(\ref{eq:heat-ocrust}).
	Indeed, the difference in $|Q_{\rm EC}(Z) - Q_{\rm EC}(Z-1)|$ calculated using the HFB-24 and HFB-27 mass model amounts to $\sim 1.3$~MeV ($\sim 0.9$~MeV) for the transition from $^{56} \mathrm{Ti}$ to $^{56} \mathrm{Ca}$ (from $^{80} \mathrm{Zn}$ to $^{80} \mathrm{Ni}$).
	
	The maximum amount of heat released per nucleon is found to be $\sim 0.1$~MeV from electron captures, and $\sim 1-2$~MeV from pycnonuclear fusions. 
    The total amount of heat deposited in the outer crust of a magnetar is comparable to that found in accreting neutron stars although the initial composition and the physical conditions are totally different~\cite{chamel2020b}. However, heat in a magnetar is deposited at higher densities, $\rho_\beta\sim 10^{10}-10^{11}$~g~cm$^{-3}$ (corresponding to pressures $P_\beta\sim 10^{29}-10^{30}$~dyn~cm$^{-2}$) for the magnetic field considered. 
	
	\subsection{Astrophysical implications}
	
	The range of densities where most of the heat from nuclear reactions is released corresponds to that determined empirically by fitting the observed thermal luminosity using magnetar cooling simulations~\cite{kaminker2006,kaminker2009}. 
	
	The time $\tau$ required for all the nuclei $(A,Z)$ to capture electrons, roughly given by \cite{chamel2021}
	\begin{equation}\label{eq:age}
		\tau  \sim  \frac{4 \pi }{B^2}\biggl[ P_\beta(A,Z,B_\star)-P_{\rm min}(A,Z,B_\star) \biggr]~\rm  Myr
	\end{equation}
	($P_{\rm min}(A,Z,B_\star)$ being the lowest pressure at which parent nuclei $(A,Z)$ were initially found), is of the same order in the different crustal layers, and more importantly of the same order as the kinematic age of magnetars (a few thousand years). 
	
	Finally, the heat power, estimated as \cite{chamel2021}
	\begin{equation}\label{eq:power}
		W^\infty\sim
		\frac{1}{\tau} \sum  \mathcal{Q}(A,Z) \mathcal{N}(A,Z)\sim 10^{35}-10^{36} \ \rm erg/s \, ,
	\end{equation}
	where $\mathcal{N}(A,Z)$ is the total number of nuclei $(A,Z)$,  
	is of the same order as that obtained in Refs.~\cite{kaminker2006,kaminker2009}. 
	
	The compression of matter caused by the spin down of the star could also trigger nuclear reactions, as studied in millisecond pulsars~\cite{IidaSato1997}. However, this heating becomes ineffective several days after the formation of the star~\cite{chamel2021} and can thus be ignored.
	
	\section{Conclusions}
	
	Our analysis suggests that electron captures by nuclei and pycnonuclear fusion reactions in the outer crust of a magnetar induced by the decay of the magnetic field could be a robust source of internal heating. Unlike the commonly accepted dissipation of elastic energy due to crust failure, the heating from electron captures remains effective even if some part of the crust is melted. Although for simplicity we have considered an extremely quantizing magnetic field, we have shown that the maximum amount of heat is essentially determined by nuclear data alone. Therefore, our estimate remains a fairly good approximation for neutron stars with lower magnetic fields, as confirmed by our recent study~\cite{ChamelFantina2022}. Only the location of the heat sources will vary substantially depending on the magnetic field strength. The models HFB-24 and HFB-27 predict similar results for the main heat sources, however significant differences are found for some minor reactions.

	\ack
	The work of N.C. was funded by Fonds de la Recherche Scientifique-FNRS (Belgium) under Grant Number IISN 4.4502.19. L.S., P.H., and J-L.Z. acknowledge the financial support from the National Science Centre (Poland) Grant Number 2018/29/B/ST9/02013. This work was also partially supported by the European Cooperation in Science and Technology Action CA16214 and the CNRS International Research Project (IRP) ``Origine des \'el\'ements lourds dans l'univers: Astres Compacts et Nucl\'eosynth\`ese (ACNu)''.

\end{document}